# Any-horizon uniform random sampling and enumeration of constrained scenarios for simulation-based formal verification

Toni Mancini, Igor Melatti, and Enrico Tronci

**Abstract**—*Model-based* approaches to the verification of non-terminating Cyber-Physical Systems (CPSs) usually rely on *numerical simulation* of the System Under Verification (SUV) model under input scenarios of possibly varying duration, chosen among those satisfying given *constraints*. Such constraints typically stem from *requirements* (or *assumptions*) on the SUV inputs and its *operational environment* as well as from the enforcement of *additional conditions* aiming at, *e.g.*, *prioritising* the (often extremely long) verification activity, by, *e.g.*, focusing on scenarios explicitly exercising *selected* requirements, or avoiding *vacuity* in their satisfaction.

In this setting, the possibility to *efficiently sample at random* (with a known distribution, *e.g.*, uniformly) within, or to efficiently *enumerate* (possibly in a uniformly random order) scenarios among those satisfying all the given constraints is a key enabler for the practical viability of the verification process, *e.g.*, via simulation-based statistical model checking.

Unfortunately, in case of non-trivial combinations of constraints, iterative approaches like Markovian random walks in the space of sequences of inputs in general *fail* in extracting scenarios according to a given distribution (*e.g.*, uniformly), and can be *very inefficient* to produce at all scenarios that are both legal (with respect to SUV assumptions) and of interest (with respect to the additional constraints). For example, in our case studies, up to 91% of the scenarios generated using such iterative approaches would need to be neglected.

In this article, we show how, given a set of constraints on the input scenarios succinctly defined by multiple *finite memory monitors*, a data structure (*scenario generator*) can be synthesised, from which *any-horizon scenarios* satisfying the input constraints can be *efficiently* extracted by (possibly uniform) random sampling or (randomised) enumeration.

Our approach enables *seamless support to virtually all simulation-based approaches to CPS verification*, ranging from simple random testing to statistical model checking and formal (*i.e.*, exhaustive) verification, when a suitable bound on the horizon or an iterative horizon enlargement strategy is defined, as in the spirit of bounded model checking.

---

## 1 INTRODUCTION

Cyber-Physical Systems (CPSs) are typically *non-terminating* systems encompassing *software* which senses and controls (in an endless loop) one or more *physical plants* (*e.g.*, motors, electrical circuits, etc.) Such systems are ubiquitous in a wide spectrum of application domains, *e.g.*, automotive, space, avionics, smart grids, systems biology, healthcare, just to mention a few.

The intrinsic complexity of industry-relevant CPSs makes their verification and certification very challenging. To this end, *model-based* approaches are widely exploited to enable *system verification* starting from the early design phases, well before an actual implementation is built (see, *e.g.*, [8]). Indeed, model-based approaches work on a *model* describing the CPS behaviour, rather than on an actual implementation.

- *Authors are with the Computer Science Department, Sapienza University of Rome, Italy.*
  *E-mail: tmancini@di.uniroma1.it, melatti@di.uniroma1.it, tronci@di.uniroma1.it*

While for, *e.g.*, digital circuits, model-based verification is usually carried out using *model-checking* techniques (see, *e.g.*, [19]), the case of CPSs is peculiar: in fact, although, from a formal point of view, they can be modelled as *hybrid systems*, model checkers for hybrid systems can only handle verification of moderately sized CPSs. Thus, *numerical simulation* is currently the main workhorse for the model-based verification of large CPSs, and is widely supported by available design tools (see, *e.g.*, Simulink, VisSim, Dymola, ESA Satellite Simulation Infrastructure SIMULUS). Such tools take as input a model of the behaviour of the CPS (for example, via systems of algebraic-differential equations plus algorithmic snippets for, *e.g.*, handling of events) along with an *operational scenario* (defining the actual system inputs and the conditions on the environment in which the CPS is operating), and provide as output the associated time course of the quantities of interest (*system trajectory*) up to a user-requested time-horizon (possibly dynamically revised during the verification process).

### 1.1 Motivation

The case of CPS models only available as *black boxes* (*e.g.*, *simulators*, our focus here) is typical for many industry-relevant systems and has far-reaching implications. In particular, (bounded-horizon) *numerical simulation* of the

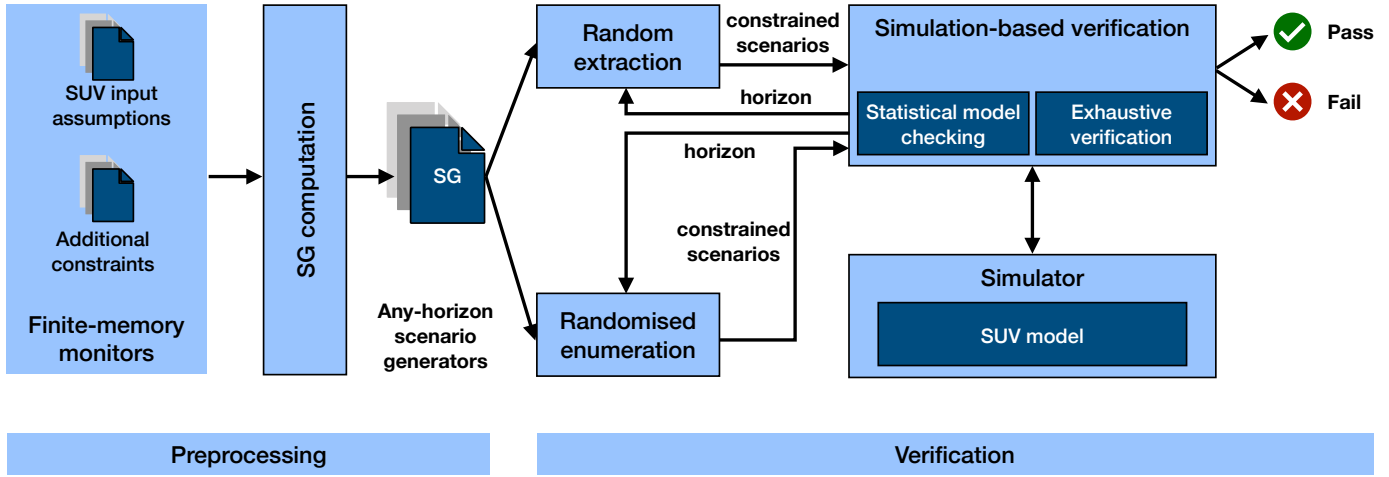

Figure 1: High-level overview of our approach.

System Under Verification (SUV) model becomes the *only viable means* to obtain the system trajectory under any given input scenario.

In this setting, carrying out a verification process of the SUV upon an *explicit, precise, and implementation-independent definition* of the set of its (non-terminating) *legal input scenarios* (which stem from requirements deriving from the SUV assumptions and constraints about its operational *environment*) is crucial for the system *certification* (for, *e.g.*, regulatory purposes). Also, the possibility to effectively *sample* (possibly *uniformly at random*) from this set or (*randomly*) enumerate it (when finite [43], [47]), is important to derive *statistical guarantees* about its correctness (*e.g.*, via Statistical Model Checking, SMC), and to spot *hard-to-find* errors (*i.e.*, errors in the system attained only under a *small* number of input scenarios).

Furthermore, *effective means to dynamically focus*, among such legal scenarios, on those that satisfy certain *additional constraints*, and to efficiently enumerate or sample at random within these subsets, enables *prioritisation* of the (often prohibitively long) verification activity, and allows one to focus on scenarios that drive the system towards specific portions of the space of behaviours, *e.g.*, to *explicitly exercise selected requirements* or to avoid *vacuity* in their satisfaction (see, *e.g.*, [12], [39], [69]).

Unfortunately, it is well known (see, *e.g.*, [17] and citations thereof) that, in case of non-trivial combinations of constraints, iterative approaches like Markovian random walks in the space of sequences of inputs in general *fail* in extracting scenarios according to a given distribution (*e.g.*, uniformly). Also, such approaches can be *very inefficient* to produce at all scenarios that are both legal (with respect to SUV assumptions and requirements on its environment) and of interest (with respect to the additional constraints). This is because the intricacies of and the inter-dependencies among requirements and additional constraints easily make such iterative approaches reach *deadlocks*. As an example, in our case studies up to 91% of random walks in the space of sequences of *legal* inputs would *fail* to produce scenarios satisfying all the provided constraints. This would translate into a *major source of inefficiency* for the verification (*e.g.*, SMC) algorithm, which would be forced to *trash out* such (possibly huge) ratio of generated random walks.

### 1.2 Contributions

In this article we rely on Finite State Machines (FSMs) as a succinct, flexible, and practical means to compositionally define (as *monitors*) requirements about legal scenarios on which the SUV shall be exercised, as well as additional constraints aiming at restricting the focus of the verification process (*e.g.*, for prioritisation needs, or to exercise some requirements avoiding vacuity in their satisfaction). Furthermore, we show how, by exploiting and combining together results from supervisory control theory and combinatorics, we can synthesise a data structure (Scenario Generator) from which any-horizon scenarios entailed by a monitor, or a combination thereof, can be efficiently extracted by (possibly uniform) random sampling or (randomised) enumeration.

Our approach enables *seamless support to virtually all simulation-based approaches to CPS verification*, ranging from simple random testing to SMC and formal (*i.e.*, exhaustive) verification, when a suitable bound on the horizon or an iterative/dynamic horizon enlargement strategy is defined, as in the spirit of bounded model checking (see, *e.g.*, [47]). Figure 1 gives a high-level overview of our approach.

### 1.3 Paper outline

This article is organised as follows. Section 2 introduces relevant background and states our formal setting. Section 3 presents our approach to the (possibly uniform) random sampling and (possibly randomised) enumeration of any-horizon constrained scenarios. Section 4 describes our three case studies and our experimental results. Section 5 is devoted to related work and Section 6 draws conclusions.

## 2 BACKGROUND AND FORMAL SETTING

In this section we introduce the relevant background notions and give the formal setting on which we build our approach.

Throughout the paper, we denote with $\mathbb{R}$, $\mathbb{R}_{0+}$ and $\mathbb{R}_+$ the sets of, respectively, all real, non-negative real, strictly positive real numbers, and with $\mathbb{N}_+$ and $\mathbb{N}$ the sets of, respectively, strictly positive and non-negative integer numbers.

We also denote with Bool = {false, true} the set of Boolean values. Furthermore, given set $A$, we denote with $A^*$ the set of infinite sequences of elements of $A$.

### Systems and System Requirements

In this article, we focus on verifying *non-terminating* systems, such as CPSs, available as black boxes via simulators. These systems are typically defined as dynamical systems [63].

*Contracts* have been widely advocated to formalise the *requirements* of a system in an implementation-independent way, in terms of *assumptions* of the inputs and *guarantees* on the outputs (see, *e.g.*, [13]). Verifying whether SUV $\mathcal{H}$ is correct with respect to its contract means to check that, whenever $\mathcal{H}$ is fed with inputs satisfying the *constraints* dictated by the contract assumptions, then it produces outputs compliant to the contract guarantees.

We assume that inputs to our SUV $\mathcal{H}$ will be given as unbounded *input time functions* in the form of time courses of *assignments* to $n$ variables $\mathbb{V} = \{v_1, v_2, \ldots, v_n\}$ (the *input variables* of $\mathcal{H}$), where each $v_i$ takes values within domain $\mathbb{U}_{v_i}$. Thus, each constraint defining the set of input scenarios to consider (*e.g.*, as dictated by the contract assumptions or by any additional constraint we might want to enforce) will be defined over (a subset of) these variables. Given any $V = \{v_{j_1}, v_{j_2}, \ldots, v_{j_k}\} \subseteq \mathbb{V}$, we denote by $\mathbb{U}_V = \mathbb{U}_{v_{j_1}} \times \cdots \times \mathbb{U}_{v_{j_k}}$ the space of assignments of variables in $V$ to values of their respective domains, and given an assignment $u \in \mathbb{U}_V$ (on some set of variables $V \subseteq \mathbb{V}$) and a subset $V'$ of $V$, we denote by $u_{|V'} \in \mathbb{U}_{V'}$ the *projection* of $u$ onto $V'$, *i.e.*, the assignment which coincides with $u$ for all variables in $V'$ and is undefined elsewhere.

### Monitors

Since our SUVs of interest are non-terminating, both assumptions and guarantees of their contracts are *time unbounded*. Hence, we need practical means to *finitely* and *conveniently* define them. A general and flexible means to define sets of unbounded time functions is via the use of *monitors*, *i.e.*, (possibly black-box) systems that scan an input time function and reject it as soon as they conclude it violates some requirement.

Given our focus on verification tasks where numerical simulation is the only means to get the trajectory of the SUV when fed with an input scenario, we assume that the SUV input space $\mathbb{U}_\mathbb{V}$ is *finite* (and, without loss of generality, *ordered*) and contract assumptions can be defined by a monitor which uses only *finite memory*.

Our assumption is in line with an engineering (rather than purely mathematical) point of view, where man-made CPSs need to satisfy the properties under verification with some degree of *robustness* with respect to the actual input time functions (see, *e.g.*, [1], [23] and references thereof).

A direct consequence of our setting is that it allows us to seamlessly support (see Section 3) the *full continuum* from contract random testing to statistical model checking up to formal (*i.e.*, exhaustive) simulation-based (robust) verification of black-box systems.

Our approach naturally applies to systems whose input values denote *events* such as faults (which can typically be finitely enumerated under a given order). However, thanks to the SUV robustness hypothesis, inputs assuming *continuous values* can be tackled by means of a suitable *discretisation* of their (ordered) domains, whilst truly *continuous-time inputs* (*e.g.*, additive noise signals) can be managed as long as they can be cast into (or suitably approximated by) *finitely parametrisable* functions, in which case the input space actually defines such a (discrete or discretised) parameter space. Practical examples of such finite parameterisations of the SUV input space (or projections thereof) are those defining limited, quantised Taylor expansions of continuous-time finite inputs, or those defining quantised values for the first coefficients (those carrying out the most information) of the Fourier series of a finite-bandwidth noise, see, *e.g.*, [1], [49]. Clearly, the most suitable approach to finitely define the SUV input space (perhaps a safe approximation thereof, consistent with the verification needs) depends on the case at hand, and should be carefully designed by the verification engineer. Our case studies in Section 4 contain uses of several of such features, and show that our setting can be easily met in practice.

Definition 1 introduces the notion of (finite-memory) monitor on top of the standard notion of FSM.

**Definition 1** (Monitor). *A monitor $\mathcal{M}$ is defined in terms of a FSM, namely a tuple $(V, X, x_0, f)$ where:*
- $V \subseteq \mathbb{V}$ *is the finite set of* input variables, *which defines the monitor input space* $\mathbb{U}_V$;
- $X$ *is the finite set of the monitor* states;
- $x_0 \in X$ *is the monitor* initial state;
- $f : X \times \mathbb{U}_V \to X$, *the monitor* transition function, *is a (possibly partial) function defining the* legal transitions *of $\mathcal{M}$.*

*We denote by Traces$(\mathcal{M}) \subseteq \mathbb{U}_V^*$ the set of* traces *of $\mathcal{M}$, that is the set of* infinite *sequences $(u_0, u_1, u_2, \ldots)$ of inputs (assignments to the input variables) associated to the infinite computation paths of $\mathcal{M}$ that start from $x_0$.*

*Given $h \in \mathbb{N}_+$, we denote by Traces$(\mathcal{M})_{|h}$ the set of* prefixes *of length $h$ of the traces of $\mathcal{M}$.*

Note that a monitor transition function $f$ can (and usually will) be partial. This allows us to easily encode within $f$ constraints on the infinite sequences of inputs entailed by the monitor (*i.e.*, the set of monitor traces).

Monitor traces can be easily interpreted as piecewise-constant discrete- or continuous-time functions. Indeed, given a time unit $\tau \in \mathbb{T} - \{0\}$ (with $\mathbb{T}$, the time-set, being $\mathbb{N}$ or $\mathbb{R}_{0+}$ for discrete- and continuous-time contracts and systems, respectively), a trace $(u_0, u_1, u_2, \ldots)$ of a monitor $\mathcal{M}$ is interpreted as time function $\mathbf{u}$ defined as $\mathbf{u}(t) = u_{\lfloor \frac{t}{\tau} \rfloor}$ for all $t \in \mathbb{T}$. This allows us, in particular, to regard traces of a monitor defining the assumptions of a contract for a system model $\mathcal{H}$ as (suitable approximations of) actual input time functions with which to feed $\mathcal{H}$ during simulation-based verification.

Monitors can be conveniently defined using a wide variety of specification languages, *e.g.*, temporal logics (see, *e.g.*, [31]), data-flow languages (see, *e.g.*, [30]), but also conventional programming and simulation languages (*e.g.*, the same language of the SUV simulator itself). Indeed, it will be enough, for our approach to work, that monitors are given as *black boxes*, in that we only need to repeatedly invoke their transition function.

Definition 2 shows that monitors can be *conjoined* by simply combining their (possibly black-box) transition functions. This will allow us to define *systems of constraints* for, *e.g.*, the whole set of assumptions of a contract starting from monitors for subspaces thereof, *e.g.*, for subsets of the requirements, plus monitors enforcing additional constraints for the scenarios of interest.

**Definition 2** (Conjoint monitor). *Let $\mathcal{M}_a = (V_a, X_a, x_{a,0}, f_a)$ and $\mathcal{M}_b = (V_b, X_b, x_{b,0}, f_b)$ be two monitors over (possibly overlapping) sets of variables.*

*The conjoint monitor $\mathcal{M} = \mathcal{M}_a \bowtie \mathcal{M}_b$ between $\mathcal{M}_a$ and $\mathcal{M}_b$ is $(V_a \cup V_b, X_a \times X_b, (x_{a,0}, x_{b,0}), f_a \bowtie f_b)$, where for all $x_a \in X_a, x_b \in X_b, u \in \mathbb{U}_{V_a \cup V_b}$:*

$$(f_a \bowtie f_b)((x_a, x_b), u) = (f_a(x_a, u_{|V_a}), f_b(x_b, u_{|V_b}))$$

*if both $f_a(x_a, u_{|V_a})$ and $f_b(x_b, u_{|V_b})$ are defined and is undefined otherwise.*

*The set of traces of $\mathcal{M}$, Traces($\mathcal{M}$), is the set of sequences $(u_0, u_1, u_2, \ldots) \in \mathbb{U}^*_{V_a \cup V_b}$ such that $(u_{0|V_a}, u_{1|V_a}, u_{2|V_a}, \ldots) \in$ Traces($\mathcal{M}_a$) and $(u_{0|V_b}, u_{1|V_b}, u_{2|V_b}, \ldots) \in$ Traces($\mathcal{M}_b$).*

Example 1 shows a contract whose assumptions are defined via a finite-memory monitor, obtained by composing monitors defining the assumptions from single sub-systems (assumption subspaces).

**Example 1.** *Consider the following contract to be implemented by the digital control system of an autopilot for a space vehicle (this is inspired from our Apollo Lunar Module Digital Autopilot, ALMA, case study in Section 4.1.3). The system receives, as an input stream from the mission-control system, requests to change the attitude of (i.e., to rotate, along one or more axes) the vehicle. Such requests are defined as assignments over input variables $V_\theta$. At each input, the new attitude is computed by the system itself on the basis of the current attitude and the new request. Such a stream of inputs is guaranteed to satisfy some additional constraints (such that, e.g., the requests can be actually obeyed by the vehicle, given its technical capabilities). All this yields monitor $\mathcal{A}_\theta$ defining legal input functions over input variables $V_\theta$. When all rotation commands and attitude values are multiple of a minimum angle and occur at time points multiple of a given time unit, monitor $\mathcal{A}_\theta$ is finite-memory.*

*The autopilot control system receives feedback about the current attitude of the vehicle from a given set of sensors, and commands a set of q on-board actuators (directional jets). Jets may be subject to temporary unavailabilities (aka* faults*) which are always recovered within a given number w of time units (t.u.) By denoting with $v_{j_i}$ the (single) input variable devoted at representing fault and repair events on the i-th jet ($i \in [1, q]$), the above yields finite-memory monitor $\mathcal{A}_{j_i}$ over the input variable $v_{j_i}$ (see Figure 2).*

*The additional requirement that the control system must be resilient to the concurrent unavailability of at most $k \leq q$ actuators yields finite-memory monitor $\mathcal{A}_{\leq k}$ over input variables $\{v_{j_1}, \ldots, v_{j_q}\}$.*

*The overall set of input variables of the contract that the autopilot must implement is thus $V = V_\theta \cup \{v_{j_1}, \ldots, v_{j_q}\}$, and the overall monitor $\mathcal{A}$ for the contract assumptions is obtained by conjoining the above sub-monitors as shown in Definition 2. Additional constraints (defined via additional monitors) on the set of scenarios of interest can be enforced similarly.*

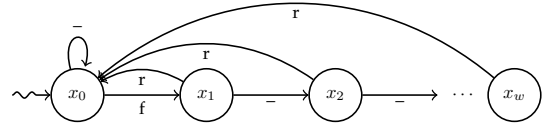

Figure 2: An explicit view of the FSM for monitor $\mathcal{A}_{j_i}$ in Example 1 ($i \in [1, q]$). Edge labels represent assignments to the single input variable $v_{j_i}$ ('–': no-op, 'f': fault, 'r': repair).

## 3 ANY-HORIZON UNIFORM RANDOM SAMPLING AND ENUMERATION OF CONSTRAINED SCENARIOS FROM MONITORS

When performing simulation-based verification of a SUV model $\mathcal{H}$, we need to simulate $\mathcal{H}$ under input scenarios satisfying all the considered requirements, *e.g.*, assumptions on the SUV inputs and any additional constraints we chose to enforce, in order to focus or prioritise the verification activity. In practice, as such scenarios are time unbounded, we need to consider suitable *finite prefixes* of them.

In this section we show how, given a monitor $\mathcal{M}$ over input variables $V$, for example obtained by conjoining the monitor defining contract assumptions and monitors defining additional constraints, we can use a combination of techniques inspired from supervisory control theory and combinatorics, in order to define two functions:

1) *nb_traces* : $\mathbb{N} \to \mathbb{N}$
2) *trace* : $\mathbb{N} \times \mathbb{N} \to \mathbb{U}^*_V$

Given a horizon $h \in \mathbb{N}$, *nb_traces($h$)* evaluates to the number of *distinct* prefixes of length $h$ of traces in *Traces($\mathcal{M}$)*; whilst, given $h \in \mathbb{N}$ and $i \in [0, nb\_traces(h) - 1]$, *trace($i, h$)* evaluates to the $i$-th (in lexicographic order) such prefix. Building on these two functions, it will be straightforward to implement any trace *random sampling policy* (possibly *uniform*), just by choosing an horizon or a range thereof (possibly dynamically during the verification activity) and then sampling over the range of indices of traces of the chosen horizon(s). *Random enumeration* of traces of a given horizon can similarly be achieved by randomly sampling a permutation of the range of indices (see, *e.g.*, [47]), and any horizon-enlargement verification policies (as in the spirit of bounded model checking [20]) can be seamlessly applied.

Functions *nb_traces* and *trace* can be implemented as to be very efficient, as it will be shown in Section 3.2. Namely, after a preliminary $\mathcal{O}\left(|\mathbb{U}_V| |X|^2\right)$ step, functions *nb_traces* and *trace* can be implemented to run in (amortised) time $\mathcal{O}(1)$ and $\mathcal{O}(h \log |\mathbb{U}_V|)$, respectively.

### 3.1 From Monitors to Scenario Generators

When monitor $\mathcal{M}$ is obtained by conjoining multiple sub-monitors, *inter-dependency* constraints may arise on the legal sequences of input values/events that make some *finite* computation paths of $\mathcal{M}$ not extensible to the infinity, and thus not defining legal prefixes of *Traces($\mathcal{M}$)*. Technically, we say that $\mathcal{M}$ might be *blocking* and, hence, may lead to *deadlocks*.

**Example 2.** *Consider again Example 1. From knowledge stemming from the root causes of the temporary unavailability of each*

*actuator (e.g., the need to charge internal capacitors or batteries after intensive use), we know that, if the attitude change requests follow certain patterns, then certain actuators (those intensively used to obey those requests) are more likely to become temporarily unavailable in the near future.*

*We might then decide to focus the verification of the autopilot system by prioritising those legal scenarios that* actually show such actuator unavailabilities. This yields monitor $\mathcal{F}$.

*Thus, our verification activity will first focus on input functions belonging to the monitor* $\mathcal{M} = \mathcal{A} \bowtie \mathcal{F}$. Monitor $\mathcal{M}$ resulting from such combination of requirements would be blocking. This is a consequence of the fact that infinite computation paths of $\mathcal{M}$ need to satisfy conflicting requirements: 'no more than $k$ actuators are simultaneously unavailable' (from $\mathcal{A}$) and 'whenever attitude change requests follows certain patterns, some given actuators will become unavailable within a certain time'. Paths not extensible to the infinity (e.g., those envisioning attitude change request patterns yielding, at some point in time, unavailabilities of more than $k$ actuators) represent scenarios that should not be considered *in the current verification stage.*

Thus, our first (and one-time preliminary) step is to *minimally restrict* monitor $\mathcal{M}$ into a new monitor $Gen(\mathcal{M})$ (Scenario Generator, SG) such that: (i) all finite paths of $Gen(\mathcal{M})$ can be extended to the infinity ($Gen(\mathcal{M})$ is *non-blocking*), and (ii) all the infinite paths of $\mathcal{M}$ are retained ($Gen(\mathcal{M})$ is *complete*).

The key concept we need for this purpose (and borrowed from supervisory control theory) is the set of *safe* states of $\mathcal{M}$ (Definition 3), *i.e.*, those states that are actually traversed by infinite paths of $\mathcal{M}$.

**Definition 3** (Safe states of a monitor). *Let* $\mathcal{M} = (V, X, x_0, f)$ *be a finite-state monitor. State* $x \in X$ *is* safe *for* $\mathcal{M}$ *if* $\Phi_f(x)$ *is true, where* $\Phi_f : X \to$ Bool *is the greatest fixed-point of the following equation:*

$$\forall x \in X \quad \Phi_f(x) \equiv \exists u, x' \quad x' = f(x, u) \wedge \Phi_f(x'). \quad (1)$$

The Scenario Generator (SG) stemming from monitor $\mathcal{M}$ (Definition 4) is obtained by restricting the transition function of $\mathcal{M}$ as to reach only safe states.

**Definition 4** (Scenario Generator). *Let* $\mathcal{M} = (V, X, x_0, f)$ *be a monitor and* $\Phi_f$ *be the function defined in Definition 3 for* $\mathcal{M}$, *with* $\Phi_f(x_0) =$ *true*.

*The Scenario Generator (SG) for* $\mathcal{M}$ *is monitor* $Gen(\mathcal{M}) = (V, X, x_0, f_{gen})$, *where, for any state* $x$ *and input* $u$, $f_{gen}(x, u) = f(x, u)$ *if* $f(x, u)$ *is defined and* $\Phi_f(f(x, u)) =$ *true, and is undefined otherwise.*

Note that Definition 4 requires that the initial state $x_0$ of $Gen(\mathcal{M})$ is a safe state for $\mathcal{M}$. Indeed, if not, then $\mathcal{M}$ entails no trace, and no SG for it exists.

A few observations are in order. The definition of $\Phi_f$ in Definition 3 is well-posed, since formula (1) is formally positive, hence, by the Knaster-Tarski Theorem [27], its greatest fixed-point exists. This implies that the SG of a monitor is unique, when it exists.

Function $\Phi_f$ for monitor $\mathcal{M} = (V, X, x_0, f)$ can be computed in time $\mathcal{O}(|\mathbb{U}_V| |X|^2)$, by starting with $\Phi_f \equiv$ true and by iteratively setting $\Phi_f(x)$ to false for those states $x$ violating (1), until the fixed-point is reached. Note that, to perform such a computation, it is enough to have $\mathcal{M}$ available as a black box, as we only need to repeatedly invoke its transition function $f$.

Finally, we point out that function $\Phi_f$ implicitly defines the (unique) most liberal *supervisory controller* for $\mathcal{M}$, *i.e.*, function $K : X \times \mathbb{U}_V \to$ Bool, such that $K(x, u) = (\exists x' \ x' = f(x, u) \wedge \Phi_f(x'))$ [71]. (Note that what are called *uncontrollable inputs* in [71] can be modelled in the starting monitor with intermediate states having only one outgoing transition.)

Proposition 1 shows that SGs satisfy our requirements (proofs of all results are available in the supplementary material).

**Proposition 1.** *Let* $\mathcal{M} = (V, X, x_0, f)$ *be a monitor and* $Gen(\mathcal{M}) = (V, X, x_0, f_{gen})$ *be its associated SG. Then:*

1) *Each finite path of* $Gen(\mathcal{M})$ *is extensible to an infinite path (non-blocking);*
2) $Gen(\mathcal{M})$ *and* $\mathcal{M}$ *have the same set of traces:* $Traces(Gen(\mathcal{M})) = Traces(\mathcal{M})$ *(completeness).*

Being based on FSMs, SGs have several desirable properties. Remark 1 states some of them.

**Remark 1.** *For every pair of monitors* $\mathcal{M}_a$ *and* $\mathcal{M}_b$:

1) $Gen(Gen(\mathcal{M}_a)) = Gen(\mathcal{M}_a)$;
2) $Gen(\mathcal{M}_a \bowtie \mathcal{M}_b) = Gen(Gen(\mathcal{M}_a) \bowtie Gen(\mathcal{M}_b))$;
3) *If* $\mathcal{M}_a$ *and* $\mathcal{M}_b$ *share no input variables (i.e., they are independent monitors), then* $Gen(\mathcal{M}_a \bowtie \mathcal{M}_b) = Gen(\mathcal{M}_a) \bowtie Gen(\mathcal{M}_b)$.

Although straightforward, properties listed in Remark 1 are very useful in practice. Namely, 2) enables the *incremental* computation of an SG defined by further constraining (*e.g.*, by imposing additional requirements on) an already computed SG (*e.g.*, an assumptions monitor). Property 3) holds when conjoining *independent* sub-monitors. This is a frequent situation occurring when defining the assumptions of a contract for a system obtained by composing monitors (on assumptions subspaces) on its components, as in Example 1. We will see in Remark 3 that in such cases, it is enough to compute the SGs of the different (typically much smaller) sub-monitors in order to extract traces of the (never computed) conjoint monitor.

### 3.2 Index-based trace extraction

Definition 5 defines two helper functions that are at the basis of the definition of functions *nb_traces* and *trace*.

**Definition 5** (Functions *ext* and $\xi$). *Let* $Gen(\mathcal{M}) = (V, X, x_0, f_{gen})$ *be the SG of* $\mathcal{M}$. *We define the following two functions:*

- *ext* $: X \times \mathbb{N} \to \mathbb{N}$, *where, for any* $x \in X$ *and* $k \in \mathbb{N}$:
  - $ext(x, k) = 1$ *if* $k = 0$;
  - $ext(x, k) = \sum_{\hat{u} \in \mathbb{U}_V} ext(f_{gen}(x, \hat{u}), k - 1)$ *otherwise.*
- $\xi : X \times \mathbb{U}_V \times \mathbb{N} \to \mathbb{N}$, *where, for any* $x \in X$ *and* $k \in \mathbb{N}$: $\xi(x, u, k) = \sum_{\hat{u} < u} ext(f_{gen}(x, \hat{u}), k)$.

**Remark 2** (Adapted from [70]). *Let* $Gen(\mathcal{M}) = (V, X, x_0, f_{gen})$ *be a* SG. *For every* $x \in X$ *and* $k \in \mathbb{N}$, *function ext(x, k) (Definition 5) evaluates to the number of all-distinct computation paths of length* $k$ *of* $Gen(\mathcal{M})$ *starting from* $x$.

Since $Gen(\mathcal{M})$ is a *deterministic* automaton, each computation path is uniquely associated to a distinct sequence of inputs. This yields our definition of function $nb\_traces : \mathbb{N} \to \mathbb{N}$ which, given a horizon $h$ returns the number $ext(x_0, h)$ of distinct prefixes of length $h$ of the traces of $Gen(\mathcal{M})$, i.e., the cardinality of $Traces(Gen(\mathcal{M}))_{|h}$. Conversely, function $\xi$, by offering a way to efficiently compute any trace prefix of length $h$ from its unique index, is at the basis of the definition of function $trace : \mathbb{N} \times \mathbb{N} \to \mathbb{U}_V^*$.

Algorithm 1 shows an implementation of functions $nb\_traces$ and $trace$ which relies on an incremental computation of maps implementing functions $ext$ and $\xi$. Provided that constant-time lookup tables (*i.e.*, *maps*) for functions $ext$ and $\xi$ are already available for all inputs up to horizon $h$, the time complexity of functions $nb\_traces$ and $trace$ is, respectively, $\mathcal{O}(1)$ and $\mathcal{O}(h \log |\mathbb{U}_V|)$. Note that, in the typical case where the verification activity asks trace prefixes of the same length $h$, computation of maps $ext$ and $\xi$ is a *one-time task*, and its cost (although in practice negligible with respect to simulation time, see Section 4) is *amortised* by the multiple (efficient) executions of function $trace$.

```
 1 global
 2   Gen(M) = (V, X, x_0, f_gen);
 3   h_max ∈ ℕ ∪ {undef}, initially undef;
 4   ext, a map of the form X × ℕ → ℕ, initially empty;
 5   ξ, a map of the form X × 𝕌_V × ℕ → ℕ, init. empty;
     // Invariant: ext(x,h) & ξ(x,u,h) defined iff h ≤ h_max
 6 function nb_traces(h)
     Input: h ∈ ℕ
 7   if h_max = undef or h > h_max then
 8     incrementally compute ext and ξ up to h;
 9     h_max ← h;
10   return ext(x_0, h);
11 function trace(i, h)
     Input: i ∈ ℕ, h ∈ ℕ
     Output: (u_0, u_1, u_2, ... u_{h-1}), i-th trace of len. h
12   if i ≥ nb_traces(h) then  error index out of bounds;
13   x ← x_0; k ← h; m ← i;
14   for j from 0 to h − 1 do
15     u_j ← max {u | ξ(x, u, k − 1) ≤ m};
16     m ← m − ξ(x, u_j, k − 1);
17     x ← f_gen(x, u_j);
18     k ← k − 1;
19   return (u_0, u_1, u_2, ... u_{h-1});
```

**Algorithm 1:** Index-based trace extraction from a SG.

**Proposition 2** (Correctness of Algorithm 1, adapted from [70]). *Functions in Algorithm 1 are such that:*

1) *For any given $h \in \mathbb{N}$, $nb\_traces(h)$ returns the cardinality of $Traces(Gen(\mathcal{M}))_{|h}$, i.e., the number of all-distinct prefixes of length $h$ of the traces of $Gen(\mathcal{M})$.*
2) *For any given $h \in \mathbb{N}$ and $i \in [0, nb\_traces(h) - 1]$, $trace(i, h)$ returns the i-th (in lexicographic order) element of $Traces(Gen(\mathcal{M}))_{|h}$.*

Remark 3 clarifies that when a monitor $\mathcal{M}$ is defined as the conjoining of two sub-monitors whose input spaces share no input variables (*independent* sub-monitors), then functions $nb\_traces$ and $trace$ for $Gen(\mathcal{M})$ can be equivalently stated in terms of the respective functions for $Gen(\mathcal{M}_a)$ and $Gen(\mathcal{M}_b)$. This allows one to avoid the computation of $\mathcal{M}$ and of $Gen(\mathcal{M})$ altogether, and to perform the index-based extraction of traces of $\mathcal{M}$ by pairing together traces extracted from (the typically much smaller) $\mathcal{M}_a$ and $\mathcal{M}_b$ (by computing only $Gen(\mathcal{M}_a)$ and $Gen(\mathcal{M}_b)$, thanks to property 3) of Remark 1).

**Remark 3.** *Let $\mathcal{M}_{ab}$ be a monitor defined as $\mathcal{M}_a \bowtie \mathcal{M}_b$, with $\mathcal{M}_a$ and $\mathcal{M}_b$ sharing no input variables.*

*Let $nb\_traces_\gamma$ and $trace_\gamma$ be the functions of Algorithm 1 for SG $Gen(\mathcal{M}_\gamma)$, with $\gamma \in \{a, b, ab\}$.*

*The following holds for all $h \in \mathbb{N}_+$:*

1) $nb\_traces_{ab}(h) = nb\_traces_a(h) * nb\_traces_b(h)$ *(where '$*$' is integer multiplication);*
2) *For all $i \in [0, nb\_traces_{ab}(h) - 1]$, $trace_{ab}(i, h) = trace_a(sel(i, a), h) \cdot trace_b(sel(i, b), h)$, where:*
   - $sel(i, a) = \left\lfloor \frac{i}{nb\_traces_b(h)} \right\rfloor$
   - $sel(i, b) = i \mod nb\_traces_b(h)$,

   *with mod being the modulo operator, and '·' the pairing of two traces: $(u_{a,0}, u_{a,1}, u_{a,2}, \ldots) \cdot (u_{b,0}, u_{b,1}, u_{b,2}, \ldots) = ((u_{a,0}, u_{b,0}), (u_{a,1}, u_{b,1}), (u_{a,2}, u_{b,2}), \ldots)$.*

## 4 EXPERIMENTAL RESULTS

In this section we define our case studies (Section 4.1) and present experimental results assessing the practical viability and scalability of our approach, focusing first on SG computation (Section 4.2) and then on index-based trace extraction from SGs (Section 4.3).

### 4.1 Case Studies

We experiment with 3 case studies consisting of contracts for the following 3 system models: Fuel Control System (FCS), Buck DC/DC Converter (BDC), Apollo Lunar Module Digital Autopilot (ALMA).

Sections 4.1.1 to 4.1.3 give more details on these systems and their associated contracts.

#### 4.1.1 Fuel Control System

The Fuel Control System (FCS) is a Simulink/Stateflow model (distributed with MathWorks Simulink) of a controller for a fault tolerant gasoline engine, which has also been used as a case study in [18], [37], [38], [42], [47], [48], [74].

The FCS has four sensors: throttle angle, speed, EGO (measuring the residual oxygen present in the exhaust gas) and MAP (manifold absolute pressure). The goal of the control system is to maintain the air-fuel ratio (the ratio between the air mass flow rate pumped from the intake manifold and the fuel mass flow rate injected at the valves) close to the stoichiometric ratio of 14.6, which represents a good compromise between power, fuel economy, and emissions. From the measurements coming from its 4 sensors, the FCS estimates the mixture ratio and provides feedback to the closed-loop control, yielding an increase or a decrease of the fuel rate.

The FCS sensors are subject to temporary faults, and the whole control system is expected to tolerate single sensor faults. In particular, if a sensor fault is detected, the FCS changes its control law and operates the engine with a higher fuel rate to compensate. In case two or more sensors fail, the FCS shuts down the engine, as the air-fuel ratio cannot be controlled.

| constraint monitor | description |
|---|---|
| 1 | Each sensor will fail every 15–20 t.u. |
| 2 | Whenever a fault on the throttle sensor occurs, a fault on the speed sensor will occur within 9–11 t.u. |
| 3 | Whenever a fault on the throttle sensor occurs, a fault on the speed sensor will occur within 13–15 t.u. |
| 4 | Whenever a fault on the throttle sensor occurs, a fault on the speed sensor will occur within 18 or 19 t.u. |
| 5 | Whenever a fault on the EGO sensor occurs, a fault on the MAP sensor will occur within 16 or 17 t.u. |
| 6 | Whenever a fault on the EGO sensor occurs, a fault on the MAP sensor will occur within 20 or 21 t.u. |

Table 1: Fuel Control System: filtering criteria.

| constraint monitor | description |
|---|---|
| 1 | $V_i$ changes at least every 6 t.u. |
| 2 | $V_i$ changes at least every 7 t.u. |
| 3 | $R$ changes at least every 5 t.u. |
| 4 | $R$ changes at least every 6 t.u. |
| 5 | $V_i$ and $R$ do not change simultaneously |
| 6 | Whenever $V_i$ changes, $R$ will change after 8 or 9 t.u. |
| 7 | Whenever $V_i$ changes, $R$ will change after 2 t.u. |

Table 2: Buck DC/DC Converter: filtering criteria.

| constraint monitor | description |
|---|---|
| 1 | Only jets number 15 and 16 may be temporarily unavailable |
| 2 | Whenever a jet is actuated for 2 consecutive t.u., it will certainly become unavailable within 3 or 4 t.u. |
| 3 | At most 1 jet is unavailable at any time |
| 4 | Rotation requests regard at most 1 axis each |
| 5 | Rotation requests regard at most 2 axes each |
| 6 | Noise signal changes for at most 1 sensor at any time |
| 7 | Noise signal for each sensor remains stable for at least 5 and at most 10 t.u. and changes by $\pm 1$ position in the given order |

Table 3: Apollo Lunar Module Digital Autopilot: filtering criteria.

The assumptions monitor $\mathcal{A}_{\text{FCS}}$ for the FCS we define for our experiments specifies that only one sensor can be faulty at any given time, and that each faulty sensor recovers within the following time bounds (in t.u.): 3–5 (throttle), 5–7 (speed), 10–15 (EGO), 13–17 (MAP). Accordingly, the *input space* of the SUV and its contract comprises 6 values: '–' (no-op), 'fault on s' (one value for each of the four sensors s) and 'repair' (which triggers repair of the single faulty sensor).

We define the additional monitors listed in Table 1 to further constrain the scenarios to focus on in various ways. In our experiments we will combine such constraints together.

### 4.1.2 Buck DC/DC Converter

The Buck DC/DC Converter (BDC) is a mixed-mode analog circuit converting the DC input voltage (denoted as $V_i$) to a desired DC output voltage ($V_o$). Buck converters are often used off-chip to scale down the typical laptop battery voltage (12–24 V) to the just few volts needed by, *e.g.*, a laptop processor (the *load*) as well as on-chip to support dynamic voltage and frequency scaling in multicore processors (see, *e.g.*, [54]). A BDC converter is *self-regulating*, *i.e.*, it is able to maintain the desired output voltage $V_o$ notwithstanding variations in the input voltage $V_i$ or in the load $R$.

A typical control system for the converter is based on software (implemented with a micro-controller) controlling a switch $u$, implemented with a MOSFET. Our case study involves the BDC software controller based on fuzzy logic originally introduced in [62].

The assumptions monitor $\mathcal{A}_{\text{BDC}}$ for the BDC we define for our experiments specifies that the input voltage $V_i$ and the load $R$ may vary during time up to at most $\pm 30\%$ of their nominal values, in steps of $\pm 5\%$ and $\pm 10\%$ of their initial values. Also, the assumptions monitor specifies that values for $V_i$ and $R$ are stable for at least 6 and 5 t.u., respectively. Finally, to have a proper set-up, $V_i$ and $R$ are assumed stable to their nominal values for the first 2 t.u. The above assumptions monitor is easily definable as $\mathcal{A}_i \bowtie \mathcal{A}_R$, where sub-monitors $\mathcal{A}_i$ and $\mathcal{A}_R$ focus on the input voltage and the load separately.

We define the additional monitors listed in Table 2 to further constrain the scenarios to focus on in various ways. In our experiments we will combine such constraints together.

### 4.1.3 Apollo Lunar Module Digital Autopilot

The Apollo Lunar Module Digital Autopilot (ALMA) is a Simulink/Stateflow model (distributed with MathWorks Simulink) defining the logic that implements the phase-plane control algorithm of the autopilot of the lunar module used in the Apollo 11 mission.

The Module is equipped with sensors (for yaw, roll and pitch) and actuators (16 reaction jets to rotate the Module along the three axes). The controller takes as input a request to change the Module attitude (*i.e.*, to perform a rotation along the three axes) as well as its current orientation as read from the sensors, and computes which reaction jets to fire to obey the request.

The assumptions monitor $\mathcal{A}_{\text{ALMA}}$ for the ALMA we define for our experiments can be defined as $\mathcal{A}_s \bowtie \mathcal{A}_{\text{rj}}$, where sub-monitor $\mathcal{A}_s$ deals with sensors noise signals, and $\mathcal{A}_{\text{rj}}$ with rotation commands and consequent actuation of reaction jets. Namely:

- $\mathcal{A}_s$ (assumptions monitor on sensors) specifies that sensors can be subject to additive noise in the form of one out of 6 predefined continuous-time signals (which, to ease application of forthcoming additional contraints, are assumed to be sorted by amplitude).
- $\mathcal{A}_{\text{rj}}$ (assumptions monitor on rotation commands and reaction jet actuations) specifies that attitude change requests do not ask the autopilot to immediately undo the rotation requested along any axis in the preceding t.u., and that the reaction jets may be temporary unavailable (unavailabilities are always recovered within 2–3 t.u.) Although $\mathcal{A}_{\text{rj}}$ can be further decomposed, for simplicity we consider it as a whole.

We define the additional monitors listed in Table 3 to further constrain the scenarios to focus on in various ways. In our experiments we will combine such constraints together.

## 4.2 Computation of Scenario Generators

In this section we show experimental results about generation of SGs associated to our case studies. Our Python/C hybrid implementation allows users to define monitors in different convenient ways: either using concise object-oriented Python code (one of the best known and simplest to use general-purpose programming languages), or via standard Functional Mock-up Unit (FMU) objects. The latter are opaque binary objects defining dynamical systems according to the Functional Mock-up Interface (FMI) open standard for model exchange. As such, FMUs can be *automatically* generated from 100+ different simulation platforms, including Modelica simulators (also open source implementations via, *e.g.*, [60]), Mathworks Stateflow/Simulink, and SBML (via, *e.g.*, the tool in [41]).

Our SG computing software expects a monitor object (either a Python object or an FMU, with other languages/formats that could be similarly supported) implementing a few API functions (mainly, a function returning the input values admissible in the current monitor state and one performing a transition from the current state given an admissible input value). Such functions can also be easily provided by the user as to define a conjoined monitor of other monitors. Our implementation computes SGs by performing a Depth-First Search (DFS) on the input monitor treated as a black box. This means it needs to access only the monitor initial state and input space, and to repeatedly invoke the monitor transition function (and get the resulting states, even if as opaque objects). Saving and restoring monitor states during search is implemented either within our software (for Python-defined monitors) or by exploiting the FMI API (for FMU-defined monitors, for which we used the implementation in [60]).

In the following experiments, we defined our monitors in Python. All computations were run on single cores of Intel(R) i7-4930K computers @ 3.40 GHz with 64 GB RAM.

Table 4 shows, for each SUV and each monitor $\mathcal{M}$ (defining contract assumptions satisfying the given filter conditions), the number of inputs of $\mathcal{M}$ (column "size of input space") as well as the time (in seconds) needed to compute $Gen(\mathcal{M})$ (column "time").

The table shows that computation of the SGs is very efficient. This is also because the decomposition properties of monitors can be often exploited (see Remarks 1 and 3). For example, the ALMA SG number 1 has been computed as a tuple of 19 sub-SGs, $(Gen_{r_1}, \ldots, Gen_{r_3}, Gen_{j_1}, \ldots, Gen_{j_{16}})$: $Gen_{r_1}, \ldots, Gen_{r_3}$ are three identical SGs associated to the assumption subspace of the rotation commands along each axis, and $Gen_{j_1}, \ldots, Gen_{j_{16}}$ are 16 identical SGs, each one associated to the assumption subspace of a single reaction jet. No combination among such 19 (independent) SGs needs to be actually computed in order to extract the associated traces (Remark 3), hence the computation time of whole SG is the overall time to compute one SG of each kind.

Clearly, when sub-monitors are defined which span multiple assumption subspaces, the SG computation may be more expensive. For example, to compute ALMA SGs number 2–6, we need to actually conjoin the 16 above single-jet SGs; and, to compute SGs number 8–9 we need to conjoin 3 identical SGs, each one defining the assumption subspace of the noise signal for each sensor, before conjoining the monitor defining constraint 6. On the other hand, the ALMA SG number 10 is never computed as a whole, but is defined as the *pair* of SGs number 5 and number 9 (hence, again its computation time is the sum of the computation times of two sub-SGs).

| SUV | SG nb. | $\mathcal{M}$ | | | $\lvert Gen(\mathcal{M}) \rvert$ |
|---|---|---|---|---|---|
| | | assumptions monitor | constraint monitors | size of input space | time [s] |
| FCS | 1 | $\mathcal{A}_{\text{FCS}}$ | – | 6 | 0.1 |
| | 2 | $\mathcal{A}_{\text{FCS}}$ | 1 | 6 | 7.99 |
| | 3 | $\mathcal{A}_{\text{FCS}}$ | 1, 3 | 6 | 4.92 |
| | 4 | $\mathcal{A}_{\text{FCS}}$ | 1, 2 | 6 | 4.61 |
| | 5 | $\mathcal{A}_{\text{FCS}}$ | 1, 4 | 6 | 6.34 |
| | 6 | $\mathcal{A}_{\text{FCS}}$ | 1, 4, 5 | 6 | 5.92 |
| | 7 | $\mathcal{A}_{\text{FCS}}$ | 1, 4, 6 | 6 | 6.55 |
| BDC | 1 | $\mathcal{A}_i$ | – | 5 | 0.19 |
| | 2 | $\mathcal{A}_R$ | – | 5 | 0.17 |
| | 3 | $\mathcal{A}_i \bowtie \mathcal{A}_R$ | – | 25 | 0.36 |
| | 4 | $\mathcal{A}_i$ | 1 | 5 | 0.12 |
| | 5 | $\mathcal{A}_i$ | 2 | 5 | 0.17 |
| | 6 | $\mathcal{A}_R$ | 3 | 5 | 0.11 |
| | 7 | $\mathcal{A}_R$ | 4 | 5 | 0.16 |
| | 8 | $\mathcal{A}_i \bowtie \mathcal{A}_R$ | 5 | 25 | 37.34 |
| | 9 | $\mathcal{A}_i \bowtie \mathcal{A}_R$ | 2, 4, 5 | 25 | 29.68 |
| | 10 | $\mathcal{A}_i \bowtie \mathcal{A}_R$ | 2, 4, 5, 6 | 25 | 1.94 |
| | 11 | $\mathcal{A}_i \bowtie \mathcal{A}_R$ | 1, 3, 5, 7 | 25 | 2.16 |
| ALMA | 1 | $\mathcal{A}_{\text{rj}}$ | – | 1 769 472 | 0.44 |
| | 2 | $\mathcal{A}_{\text{rj}}$ | 1 | 108 | 0.44 |
| | 3 | $\mathcal{A}_{\text{rj}}$ | 1, 2 | 108 | 448.88 |
| | 4 | $\mathcal{A}_{\text{rj}}$ | 1, 2, 3 | 108 | 247.27 |
| | 5 | $\mathcal{A}_{\text{rj}}$ | 1, 2, 3, 4 | 108 | 55.19 |
| | 6 | $\mathcal{A}_{\text{rj}}$ | 1, 2, 3, 5 | 108 | 188.3 |
| | 7 | $\mathcal{A}_s$ | – | 27 | 2.94 |
| | 8 | $\mathcal{A}_s$ | 6 | 27 | 1.33 |
| | 9 | $\mathcal{A}_s$ | 6, 7 | 27 | 782.2 |
| | 10 | $\mathcal{A}_{\text{ALMA}}$ | 1, 2, 3, 4, 6, 7 | 2916 | 837.39 |

Table 4: Computation times of Scenario Generators (SGs).

Overall, Table 4 shows that, even when we need to conjoin multiple sub-SGs because of the presence of constraint monitors spanning several assumption subspaces, *the overall computation times are negligible* when compared with the time needed to perform any kind of simulation-based verification of the SUV. As an example, the time to compute the most expensive SG in Table 4 (number 10) equals the time to simulate just a few input traces of the Simulink ALMA SUV model (*e.g.*, less than 50 traces for 200 t.u. each, since simulating each of them takes around 20 seconds).

## 4.3 Index-based trace extraction from Scenario Generators

The decomposed representation of an SG as a tuple of sub-SGs is also exploited when extracting traces, by relying on the equivalences of Remark 3 when applicable. Figure 3 shows the efficiency and scalability of our monitor-based approach to scenario generation for simulation-based verification of contracts for our CPSs. Namely, for each SG $Gen(\mathcal{M})$ reported in Table 4 and for different values for the time horizon $h$, the following statistics are plotted.

**Number of traces.** This is the overall number of traces of length $h$ entailed by $Gen(\mathcal{M})$, as returned by function nb_traces of Algorithm 1. Unsurprisingly, especially for SGs defined as tuples of many sub-SGs, such numbers can easy jump to tremendously high values.

**Trace extraction time.** This is the average time (in seconds) for the extraction of a single trace from its unique index, *i.e.*, the average execution time of function *trace* of Algorithm 1. Average trace extraction times have been computed by extracting 1000 same-horizon traces *uniformly at random*, and *include* the (amortised) time to compute maps *ext* and $\xi$ (whose share essentially represents $\sim 100\%$ of the overall time). In case of SGs defined as tuples of sub-SGs, the amortised time to extract a single trace of horizon $h$ is the sum of the times to extract a single trace of horizon $h$ from each sub-SG. In all cases, the average trace extraction times are *negligible* (typically a tiny fraction of a second).

**Selectivity of constraint monitors.** This is the ratio between the number of traces of length $h$ entailed by $Gen(\mathcal{M})$ and the number of traces (of the same horizon) entailed by the SG computed by *ignoring* any constraint monitor. This measures the selectivity of the (conjoined) applied constraint monitors, *i.e.*, their power to narrow down the set of traces of interest for the current verification stage. The *extremely small ratios* show the crucial importance of constraint monitors to restrict the set of scenarios of interest to those satisfying additional requirements, and to obtain more reasonable numbers of traces on which to temporarily focus the verification activity. By defining a *priority list* of such additional requirements, the verification activity itself can be guided in a well-controlled way towards the scenarios of interest.

**SG selectivity (comparison against baseline).** This statistic directly measures the benefits of computing the SG from the starting monitor $\mathcal{M}$ (or collection of monitors) *before* extracting traces. Namely, SG selectivity is the ratio between the number of traces of length $h$ entailed by $Gen(\mathcal{M})$ and the number of traces (of the same horizon) entailed by the (possibly blocking) monitor $\mathcal{M}$. Such ratio can be *very small* (as small as 9% in our experiments), especially when combinations of constraints are involved. Indeed, by skipping the construction of the SG, the verification process (*e.g.*, a statistical model checker sampling Markovian random walks directly on $\mathcal{M}$, which can be thought of as a baseline) could possibly need, on our examples, to trash out as many as 91% of the sampled traces (for the FCS case study), once discovering that they do not represent bounded-horizon prefixes of legal infinite traces. Clearly, such ratio can be made arbitrarily small by using more intricate constraints.

## 5 RELATED WORK

Several approaches, both explicit (see, *e.g.*, [25], [70]) and symbolic (see, *e.g.*, [2], [17]) have been proposed to sample uniformly at random finite paths from a finite-state combinatorial structure (automaton, combinatorial circuit, or SAT instance). Differently from such methods, since we focus on *finite prefixes of infinite paths*, we first need to restrict the input monitor (a FSM or a collection thereof) to its non-blocking portion, via the concept of Scenario Generator, by essentially computing the most liberal supervisory controller of it.

Automatic generation of admissible operational scenarios to support simulation-based verification of CPSs has been investigated in [42], [44], [45], [46], [48] as for fully general Simulink models, and in [67] as for ESA SIMSAT models. We note however that the above approaches need to generate *upfront* all scenarios of a given length satisfying the assumptions of the SUV contract in order to be able to select a subset thereof at random. When the number of the admissible scenarios is very large (our case) this can be impossible. Also, in case of blocking combinations of assumptions and filtering condition, each generated scenario should be checked after being generated to avoid to waste time to simulate the SUV model on not-truly-admissible scenarios. As shown in Section 4, this may cause major loss of efficiency of the verification process. The approach proposed in this article solves this problem, by delivering a scenario generator that can efficiently extract scenarios satisfying the SUV contract assumptions (plus, possibly, additional constraints) of *any* length (possibly dynamically revised during verification) in *any given* order (*e.g.*, uniformly at random, or as a randomised enumeration). This, in turn, enables an *embarrassingly parallel* approach to SUV contract verification.

System Level Formal Verification (SLFV) of CPS via simulation-based bounded model checking has been studied in many contexts. Here are a few examples. Simulation-based reachability analysis for large linear continuous-time dynamical systems has been investigated in [9]. A simulation based data-driven approach to verification of hybrid control systems described by a combination of a black-box simulator for trajectories and a white-box transition graph specifying mode switches has been investigated in [24]. Formal verification of discrete time Simulink models (*e.g.*, Stateflow or models restricted to discrete time operators) with small domain variables has been investigated in, *e.g.*, [15], [55], [64], [68]. We note, however, that none of the above approaches supports simulation-based bounded model checking of arbitrary simulation models, starting from contracts defined via monitors.

Simulation-based approaches to SMC have been also widely investigated, *e.g.*, [11], [14], [16], [26], [28], [29], [35], [36], [72], [73]. Simulink models for CPS have been studied in [21], [74], mixed-analog circuits have been analysed in [18]. Smart grid control policies have been considered in [32], [50], [51], [52], biological models have been studied in [53], [56], [61], [66]. Our scenario generators are independent of the SMC technique employed and can be seamlessly applied to any of them. Also, thanks to the possibility to sample the set of scenarios *without* replacement, our algorithm (differently than Monte Carlo–based SMC) is able to answer a contract verification problem with *certainty* in those cases where a not-too-large number of bounded-horizon scenarios is enough to consider, and enough time is allocated to simulate all of them (see, *e.g.*, [43], [47]).

Parallel approaches to SMC have been investigated, *e.g.*, in [5] in the context of probabilistic properties. Our approach enables *embarrassingly parallel* verification. This is because, once a SG has been generated and a verification horizon has been selected, the range of indices of admissible traces can be split upfront into subintervals, and independent verification processes can be run in parallel on each of them.

Simulation-based *falsification* of CPS properties has been extensively investigated for Simulink models. Examples are in [1], [3], [6], [33], [59]. In particular, [1] falsifies a metric temporal logic property through Monte Carlo optimisation guided by a robustness metric defined by the property. Although [1] does not provide a quantification of the level of assurance achieved when no counterexample is found (as instead SMC does), we note that the search strategy in [1] can falsify properties that would take too long to falsify using uniform sampling. Developing approaches that can provide the benefits of both approaches is, to the best of our knowledge, an open problem and an important direction for future work.

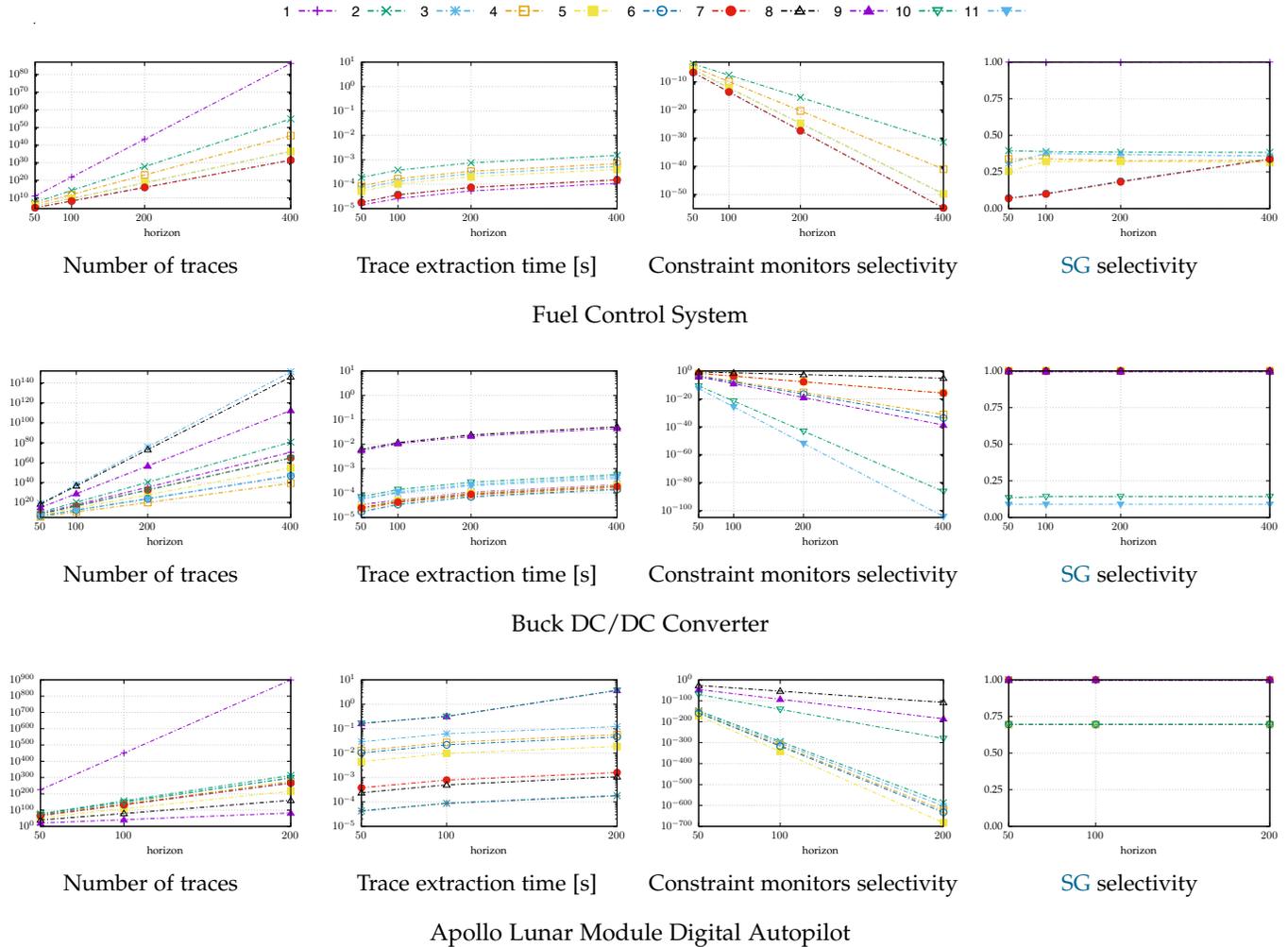

Figure 3: Number of entailed traces, amortised trace extraction time (over 1000 traces), constraint monitor as well as SG selectivity (showing savings with respect to a Markovian random walk generator in the input monitors, our baseline) for different horizons (one curve per SG of Table 4).

The supervisory control problem has been introduced in [57], [58] in a language-theoretic setting. The work in [10] (respectively, [65]) presents a symbolic (OBDD-based) algorithm for the synthesis of (optimal) supervisory controllers for finite state systems. Since then, supervisory control theory has been used in many settings, *e.g.*, analysis of database transaction execution [40], concurrent programs synthesis [34], confidentiality-enforcing in security [22], model-based system engineering [7], synthesis of control software [4], [54]. An up to date overview is in [71]. In this article, we use supervisory control theory through an *explicit* (since we face with possibly black-box monitors) implementation of the symbolic algorithm in [10].

## 6 CONCLUSIONS

In this article we focused on the (possibly uniform) random sampling and (randomised) enumeration of *constrained* input scenarios of *any horizon* for the simulation-based verification of non-terminating Cyber-Physical Systems (CPSs).

By relying on finite state machines as a succinct, flexible, and practical means to define *constraints* to be satisfied by the System Under Verification (SUV) input scenarios (where such constraints stem from *requirements* on the SUV inputs, *e.g.*, assumptions and a definition of its *operational environment*, as well as from *additional conditions* aiming at dynamically *restricting* the focus of the verification process, *e.g.*, to exercise some requirements or to counteract *vacuity* in their satisfaction), we showed how to exploit and combine together results from supervisory control theory and combinatorics in order to synthesise a data structure (Scenario Generator, SG) from which *any-horizon* scenarios can be efficiently *sampled* (possibly *uniformly* at random) or (randomly) *enumerated*.

Our approach can be seamlessly exploited in *virtually all simulation-based approaches to CPS verification*, ranging from simple random testing to statistical model checking and formal (*i.e.*, exhaustive) verification (*e.g.*, bounded model checking).

### Acknowledgements

This work was partially supported by: Italian Ministry of University & Research under grant "Dipartimenti di eccellenza 2018–2022" of the Dept. Computer Science, Sapienza Univ. of Rome; INdAM "GNCS Project 2020"; Sapienza U. projects RG11816436BD4F21, RG11916B892E54DB, RP11916B8665242F; Lazio POR FESR projects E84G20000150006, F83G17000830007. Authors are grateful to: the anonymous reviewers for their comments; Michele Laurenti, who developed an initial prototype of our FMU-based SG computing software as part of his B.Sc. thesis; Vadim Alimguzhin, for discussions and technical assistance on our implementation.

## APPENDIX

In this section we show proofs of all results stated in the article. To ease reading, proofs are presented in a detailed step-by-step style. Definition and equation numbers occurring here refer to the main article.

**Proposition 1.** Let $\mathcal{M} = (V, X, x_0, f)$ be a monitor and $Gen(\mathcal{M}) = (V, X, x_0, f_{\text{gen}})$ be its associated SG. Then:

1) Each finite path of $Gen(\mathcal{M})$ is extensible to an infinite path (non-blocking);
2) $Gen(\mathcal{M})$ and $\mathcal{M}$ have the same set of traces: $Traces(Gen(\mathcal{M})) = Traces(\mathcal{M})$ (completeness).

*Proof.*

**Point 1)** This point follows directly from the definition of the set of safe states of $\mathcal{M}$ (Definition 3) and of the transition function of $Gen(\mathcal{M})$ (Definition 4).

**Point 2)** Since the transition function of $Gen(\mathcal{M})$ is a restriction of the transition function of $\mathcal{M}$, infinite computation paths of $Gen(\mathcal{M})$ are also infinite computation paths of $\mathcal{M}$.

We thus need to show that each infinite computation paths of $\mathcal{M}$ is indeed also an infinite computation path of the restricted automaton $Gen(\mathcal{M})$. The proof is as follows.

1) Assume, for the sake of contradiction, that there exists an infinite path $(x_0, u_0, x_1, u_1, x_2, u_2, \ldots)$ of $\mathcal{M}$ that is not a path for $Gen(\mathcal{M})$.
2) This means that there must be a state, let $x_i$ ($i > 0$) be the first one, such that $f(x_{i-1}, u_{i-1}) = x_i$ but $f_{\text{gen}}(x_{i-1}, u_{i-1})$ is undefined.
3) By Definition 4, $x_i$ must be such that $\neg\Phi_f(x_i)$.
4) By (1), this means that $\neg\exists u, x \;\; x = f(x_i, u) \land \Phi_f(x)$.
5) As a special case (when $u = u_i$ and $x = x_{i+1}$), $x_{i+1} = f(x_i, u_i)$ must be such that $\neg\Phi_f(x_{i+1})$.
6) By iterating, we have $\neg\Phi_f(x)$ for all $x \in \{x_i, x_{i+1}, \ldots\}$ (where this set is finite since $X$ is finite).
7) Now consider function $\Phi'_f : X \to$ Bool defined as $\Phi'_f(x) = $ true if $\Phi_f(x)$ or $x \in \{x_i, x_{i+1}, \ldots\}$, and false otherwise.
8) Since function $\Phi'_f$ satisfies (1) and $\{x \mid \Phi_f(x)\} \subset \{x \mid \Phi'_f(x)\}$, $\Phi_f$ cannot be the greatest fixed-point of equation (1). Contradiction.

□

**Lemma 1.** *For every pair of monitors $\mathcal{M}_a = (V_a, X_a, x_{a,0}, f_a)$ and $\mathcal{M}_b = (V_b, X_b, x_{b,0}, f_b)$ it holds:*

$$\forall x_a, x_b, u \;\; \Phi_{(f_a \bowtie f_b)}([f_a \bowtie f_b]((x_a, x_b), u)) \to \Phi_{f_a}(f_a(x_a, u_{|V_a})) \land \Phi_{f_b}(f_b(x_b, u_{|V_b})).$$

*Proof.*

1) Let $x_a, x_b, u$ such to make the antecedent true.
2) Then, $[f_a \bowtie f_b]((x_a, x_b), u)$ is defined.
3) By Definition 2, $f_\gamma(x_\gamma, u_{|V_\gamma})$ is defined for all $\gamma \in \{a, b\}$.
4) Assume, by contradiction, that $\neg\Phi_{f_\gamma}(f_\gamma(x_\gamma, u_{|V_\gamma}))$, for some $\gamma \in \{a, b\}$.
5) Then, by Definition 4, $f_{\gamma,\text{gen}}(x_\gamma, u_{|V_\gamma})$ would be undefined.
6) By Definition 2, $[f_a \bowtie f_b]_{\text{gen}}((x_a, x_b), u)$ would be undefined as well.
7) Since we know (item 2 above) that $[f_a \bowtie f_b]((x_a, x_b), u)$ is defined, by Definition 4 it must be $\neg\Phi_{(f_a \bowtie f_b)}([f_a \bowtie f_b]((x_a, x_b), u))$. Contradiction.

□

**Remark 1.** *For every pair of monitors $\mathcal{M}_a$ and $\mathcal{M}_b$:*

1) $Gen(Gen(\mathcal{M}_a)) = Gen(\mathcal{M}_a)$;
2) $Gen(\mathcal{M}_a \bowtie \mathcal{M}_b) = Gen(Gen(\mathcal{M}_a) \bowtie Gen(\mathcal{M}_b))$;
3) *If $\mathcal{M}_a$ and $\mathcal{M}_b$ share no input variables (i.e., they are independent monitors), then $Gen(\mathcal{M}_a \bowtie \mathcal{M}_b) = Gen(\mathcal{M}_a) \bowtie Gen(\mathcal{M}_b)$.*

*Proof.* Let $\mathcal{M}_a = (V_a, X_a, x_{a,0}, f_a)$ and $\mathcal{M}_b = (V_b, X_b, x_{b,0}, f_b)$.

**Property 1)** From Definition 3 and **??**, it immediately follows that the transition function of $Gen(Gen(\mathcal{M}_a))$ is equal to that of $Gen(\mathcal{M}_a)$.

**Property 2)** We will show below that $\Phi_{(f_a \bowtie f_b)} \equiv \Phi_{(f_{a,\text{gen}} \bowtie f_{b,\text{gen}})}$. From Definition 4, this implies that the transition functions of the two SGs on both sides of the equation are equal.

In the following, $x = (x_a, x_b) \in X = X_a \times X_b$ and $u \in \mathbb{U}_{V_a \cup V_b}$.

**Proof of** $\forall x \;\; \Phi_{(f_{a,\text{gen}} \bowtie f_{b,\text{gen}})}(x) \to \Phi_{(f_a \bowtie f_b)}(x)$:

1) Let $x$ be any state such that $\Phi_{(f_{a,\text{gen}} \bowtie f_{b,\text{gen}})}(x)$ is true.
2) Then, by Definition 3, exists $u$ such that $\Phi_{(f_{a,\text{gen}} \bowtie f_{b,\text{gen}})}([f_{a,\text{gen}} \bowtie f_{b,\text{gen}}](x, u))$ is true.
3) Since $(f_{a,\text{gen}} \bowtie f_{b,\text{gen}})$ is a restriction of $(f_a \bowtie f_b)$, also $\Phi_{(f_a \bowtie f_b)}([f_a \bowtie f_b](x, u))$ must be true.

**Proof of** $\forall x \;\; \Phi_{(f_a \bowtie f_b)}(x) \to \Phi_{(f_{a,\text{gen}} \bowtie f_{b,\text{gen}})}(x)$:

1) Let $x = (x_a, x_b)$ be any state such that $\Phi_{(f_a \bowtie f_b)}(x)$ is true.
2) By Definition 3, there must exist $u$ such that $\Phi_{(f_a \bowtie f_b)}([f_a \bowtie f_b](x, u))$ is true.
3) Let $\overline{u}$ one such $u$. By Lemma 1, for all $\gamma \in \{a, b\}$ we have: $\Phi_{f_\gamma}(f_\gamma(x_\gamma, \overline{u}_{|V_\gamma}))$.

4) Hence, $f_{\gamma,\text{gen}}(x_\gamma, \overline{u}_{|V_\gamma})$ is defined for all $\gamma$, and thus $[f_{a,\text{gen}} \bowtie f_{b,\text{gen}}](x, \overline{u})$ is defined as well (Definition 2).

5) By Definition 3, the above implies that $\Phi_{(f_{a,\text{gen}} \bowtie f_{b,\text{gen}})}(x)$ is true (with $\overline{u}$ being a witness for variable $u$ in (1)).

**Property 3)** It will be enough to show that, when $V_a \cap V_b = \emptyset$, $Gen(Gen(\mathcal{M}_a) \bowtie Gen(\mathcal{M}_b)) = Gen(\mathcal{M}_a) \bowtie Gen(\mathcal{M}_b)$ and then apply property 2).

1) Let $f = f_{a,\text{gen}} \bowtie f_{b,\text{gen}}$ the transition function of $Gen(\mathcal{M}_a) \bowtie Gen(\mathcal{M}_b)$ (right side of the above equation).

2) By Definition 4, the transition function of the SG on the left side is $f_{\text{gen}}$, defined as: $\forall x = (x_a, x_b), u$ $f_{\text{gen}}(x,u) = f(x,u)$ if $\Phi_f(f(x,u))$ and undefined otherwise.

3) We need to show that $f_{\text{gen}} \equiv f$. By the previous point, it suffices to show that $\Phi_f(f(x,u))$ is true whenever $f(x,u)$ is defined.

4) Let $x = (x_a, x_b)$ and $u = (u_a, u_b)$ such that $f(x,u) = (f_{a,\text{gen}}(x_a, u_a), f_{b,\text{gen}}(x_b, u_b))$ is defined. This means that, for all $\gamma \in \{a, b\}$, $f_{\gamma,\text{gen}}(x_\gamma, u_\gamma)$ is defined.

5) Since $f_{\gamma,\text{gen}}$ ($\gamma \in \{a,b\}$) is the transition function of an SG, by Definition 4, we have that $\Phi_{f_{\gamma,\text{gen}}}(f_{\gamma,\text{gen}}(x_\gamma, u_\gamma))$ is true, which means (Definition 3) that $\Phi_{f_{\gamma,\text{gen}}}(x_\gamma)$ is true (with $u_\gamma$ being a witness for variable $u$ in (1)).

6) This implies that $\Phi_f(f(x,u)) = \Phi_f(f((x_a, x_b), (u_a, u_b))) = \Phi_f((f_{a,\text{gen}}(x_a, u_a), f_{b,\text{gen}}(x_b, u_b))) = \text{true}$, because otherwise, with a reasoning analogous to that in the proof of property 2) of Proposition 1, it would not be the greatest fixed-point.

$\square$

**Remark 2** (Adapted from [70]). Let $Gen(\mathcal{M}) = (V, X, x_0, f_{\text{gen}})$ be a SG. For every $x \in X$ and $k \in \mathbb{N}$, function $ext(x,k)$ (Definition 5) evaluates to the number of all-distinct computation paths of length $k$ of $Gen(\mathcal{M})$ starting from $x$.

*Proof.* See discussion on formula (2) in [70]. $\square$

**Proposition 2** (Correctness of Algorithm 1, adapted from [70]). Functions in Algorithm 1 are such that:

1) For any given $h \in \mathbb{N}$, $nb\_traces(h)$ returns the cardinality of $Traces(Gen(\mathcal{M}))_{|h}$, i.e., the number of all-distinct prefixes of length $h$ of the traces of $Gen(\mathcal{M})$.

2) For any given $h \in \mathbb{N}$ and $i \in [0, nb\_traces(h) - 1]$, $trace(i,h)$ returns the $i$-th (in lexicographic order) element of $Traces(Gen(\mathcal{M}))_{|h}$.

*Proof.* Point 1) follows directly from Remark 2. For point 2), see algorithm "unrank" in [70]. $\square$

**Remark 3.** Let $\mathcal{M}_{ab}$ be a monitor defined as $\mathcal{M}_a \bowtie \mathcal{M}_b$, with $\mathcal{M}_a$ and $\mathcal{M}_b$ sharing no input variables.

Let $nb\_traces_\gamma$ and $trace_\gamma$ be the functions of Algorithm 1 for SG $Gen(\mathcal{M}_\gamma)$, with $\gamma \in \{a, b, ab\}$.

The following holds for all $h \in \mathbb{N}_+$:

1) $nb\_traces_{ab}(h) = nb\_traces_a(h) * nb\_traces_b(h)$ (where '$*$' is integer multiplication);

2) For all $i \in [0, nb\_traces_{ab}(h) - 1]$, $trace_{ab}(i,h) = trace_a(sel(i,a), h) \cdot trace_b(sel(i,b), h)$, where:
   - $sel(i,a) = \left\lfloor \frac{i}{nb\_traces_b(h)} \right\rfloor$
   - $sel(i,b) = i \bmod nb\_traces_b(h)$,
   
   with mod being the modulo operator, and '$\cdot$' the *pairing* of two traces: $(u_{a,0}, u_{a,1}, u_{a,2}, \ldots) \cdot (u_{b,0}, u_{b,1}, u_{b,2}, \ldots) = ((u_{a,0}, u_{b,0}), (u_{a,1}, u_{b,1}), (u_{a,2}, u_{b,2}), \ldots)$.

*Proof.*

**Point 1)**

1) By property 3) of Remark 1, we have that $Gen(\mathcal{M}_{ab}) = Gen(\mathcal{M}_a) \bowtie Gen(\mathcal{M}_b)$.

2) Since $Gen(\mathcal{M}_a)$ and $Gen(\mathcal{M}_b)$ have no variables in common, Definition 2 ensures that the pairing $((u_{a,0}, u_{b,0}), (u_{a,1}, u_{b,1}), (u_{a,2}, u_{b,2}), \ldots)$ between each trace $(u_{a,0}, u_{a,1}, u_{a,2}, \ldots)$ of $Gen(\mathcal{M}_a)$ and each trace $(u_{b,0}, u_{b,1}, u_{b,2}, \ldots)$ of $Gen(\mathcal{M}_b)$ yields a trace of $Gen(\mathcal{M}_{ab})$.

3) The above yields $nb\_traces_{ab}(h) \geq nb\_traces_a(h) * nb\_traces_b(h)$ for any horizon $h$.

4) The converse is always true by Definition 2.

**Point 2)**

1) For any horizon $h$ and any index $i \in [0, nb\_traces_{ab}(h) - 1]$, from point 1) we have $sel(i,a) \in [0, nb\_traces_a(h) - 1]$ $sel(i,b) \in [0, nb\_traces_b(h) - 1]$. Hence, the two traces selected from $Gen(a)$ and $Gen(b)$ do exist.

2) Also, the fact that, for each $0 \leq i_1 < i_2 < nb\_traces_{ab}(h)$, $trace_a(sel(i_1, a), h) \cdot trace_b(sel(i_1, b), h)$ is lexicographically lower than $trace_a(sel(i_2, a), h) \cdot trace_b(sel(i_2, b), h)$ comes straightforwardly from the following observations:

   a) $sel(i_1, a) \leq sel(i_2, a)$, hence, by Proposition 2, $trace_a(sel(i_1, a), h)$ is lexicographically lower than or equal to $trace_a(sel(i_2, a), h)$;
   
   b) if $sel(i_1, a) = sel(i_2, a)$, then $sel(i_1, b) < sel(i_2, b)$, thus, again, $trace_b(sel(i_1, b), h)$ is lexicographically lower than $trace_b(sel(i_2, b), h)$;
   
   c) the definition of the pairing operator '$\cdot$'.

$\square$